\documentclass[12pt, reqno]{amsart}

\usepackage[dcucite]{harvard}
\usepackage{amsmath,amssymb,harvard}
\long\def\comment#1{}
\long\def\comment#1{}
\newtheorem{theorem}{Theorem}

\theoremstyle{definition}

\newtheorem{remark}{Comment}[section]
\numberwithin{remark}{section}
\newtheorem{example}{Example}
\newcommand{\citen}{\citeasnoun}

\newcommand{\be}{\begin{eqnarray}}
\newcommand{\ee}{\end{eqnarray}}

\newcommand{\ba}{\begin{array}}
\newcommand{\ea}{\end{array}}
\newcommand{\bs}{\begin{align}\begin{split}\nonumber}
\newcommand{\bsnumber}{\begin{align}\begin{split}}
\newcommand{\es}{\end{split}\end{align}}

\newcommand{\sss}{\scriptscriptstyle}

\renewcommand{\(}{\left(}
\renewcommand{\)}{\right)}
\renewcommand{\[}{\left[}
\renewcommand{\]}{\right]}

\renewcommand{\leq}{\leqslant}
\renewcommand{\geq}{\geqslant}

\begin{document}
\title[IVQR] {Quantile Models with Endogeneity}
\author[Chernozhukov \ Hansen]{V. Chernozhukov \and C. Hansen}

\date{First version:  September 2011,  this version \today.  We would like to thank the editor, Isaiah Andrews, Denis Chetverikov, and Ye Luo  for excellent comments and much help.}

\begin{abstract}  In this article, we review quantile models with endogeneity.  We focus on models that achieve identification through the use of instrumental variables and discuss conditions under which partial and point identification are obtained.  We discuss key conditions, which include monotonicity and full-rank-type conditions, in detail.  In providing this review, we update the identification results of \citen{iqr:ema}.  We illustrate the modeling assumptions through economically motivated examples.  We also briefly review the literature on estimation and inference.

Key Words:  identification, treatment effects, structural models, instrumental variables
\end{abstract}

\maketitle

\section{Introduction}

Quantile regression is a tool for estimating
conditional quantile models that has been used in many empirical
studies and has been studied extensively in theoretical econometrics; see \citen{koenker:1978} and \citen{koenker:book}. One of quantile regression's most appealing
features is its ability to estimate quantile-specific effects that
describe the impact of covariates not only on the center but also
on the tails of the conditional outcome distribution. While the central
effects, such as the mean effect obtained through conditional mean
regression, provide interesting summary statistics of the impact
of a covariate, they fail to describe the full distributional
impact unless the conditioning variables affect the central and the tail
quantiles in the same way. In addition, researchers are interested in the
impact of covariates on points other than the center of the
conditional distribution in many cases. For example, in a study of the
effectiveness of a job training program, the effect of training on
the lower tail of the earnings distribution conditional on worker characteristics may be of
more interest than the effect of training on the
mean of the distribution.

In observational studies, the variables of
interest (e.g. education or prices) are often endogenous.  Just as with the conventional linear model, endogeneity of covariates renders the conventional quantile regression inconsistent for estimating the causal (structural) effects of covariates on
the quantiles of economic outcomes.  One approach to addressing this
problem is to generalize the instrumental variables framework to allow for estimation of quantile models.
In this paper, we review developments in instrumental variables approaches to modeling and estimating quantile treatment (structural) effects (QTE) in the presence of endogeneity.

We focus our review on the modeling framework of \citen{iqr:ema} which provides
conditions for identification of the QTE without functional form
assumptions. The principal identifying assumption of the model is the imposition
of conditions which restrict how rank variables (structural errors) may vary
across treatment states. These conditions allow the use of instrumental variables to overcome the
endogeneity problem and recover the true QTE.  This framework also ties naturally to
simultaneous equations models, corresponding to a  structural simultaneous equation model
with non-additive errors.
Within this framework, estimation and inference procedures for linear quantile models have been developed by \citen{iqr:joe}, \citen{iqr:joe2}, \citen{fsqr}, and \citen{JunWeakIdIVQR}; nonparametric estimation has been considered by \citen{CIN:JoE},
\citen{HorowitzLeeNonparametricIVQR}, and \citen{GagliardiniScaillet}; and inference with discrete outcomes has been explored by \citen{ChesherDiscrete}.
Moreover, the modeling framework provides a foundation for other estimation methods based on IV
median-independence and more general quantile-independence conditions as in \citen{abadie:1997}, \citen{mcmc}, \citen{chen:linton},
\citen{hong:tamer}, \citen{honore:hu}, and \citen{sakata}.  It is also important to note that the modeling framework we review can be used to study nonparametric identification of structural economic models in cases where quantile effects are not necessarily the chief objects of interest.  \citen{BerryHaileDiscreteChoiceId} provide an excellent example of this in the context of discrete choice models with endogeneity.

We also briefly review other modeling approaches for quantile effects with endogenous covariates.
\citen{abadie} consider a QTE model for the sub-population of ``compliers" which applies to
binary endogenous variables with binary instruments. \citen{imbens:newey}, \citen{chesher}, \citen{SLeeTriangularIVQR}, and \citen{koenker:ma}  use models with
triangular structures and show how control functions can be constructed and used to
estimate structural objects of interest.  While these models share some features with the model of \citen{iqr:ema}, the three approaches are non-nested in general.

Quantile models with endogeneity have been used in many empirical studies in economics.  See \citen{abadie}; \citen{401k}; \citen{hausman:sidak}; \citen{IVQRApp:AirTraffic}; \citen{IVQRApp:Union}; \citen{IVQRApp:Agriculture};
\citen{IVQRApp:Insurance};  \citen{IVQRApp:BirthWeight}; \citen{IVQRApp:Columbia}; \citen{IVQRApp:Autor}; and \citen{IVQRApp:Somainiy} among others.  We do not provide a review of empirical applications but note these papers  provide further discussion of how the instrumental variables quantile model relates to their specific framework and illustrate some of the rich effects that one can estimate using quantile methods.

\section{An IV Quantile Model}

In this section, we present an instrumental variable model for quantile treatment effects (QTE), its main econometric
implication, and the principal identification result.

\subsection{Framework} Our model is developed within the conventional potential
(latent) outcome framework, e.g. \citen{heckman:robb}.  Potential
real-valued outcomes which vary among individuals or
observational units are indexed against potential treatment
states $d \in \mathcal{D}$ and denoted $Y_d$. The potential outcomes
$\{ Y_{d}\}$ are latent because, given the selected treatment
$D$, the observed outcome for each individual or observational
unit is only one component
$$ Y := Y_{D}$$ of the potential outcomes vector $\{Y_d\}$.
Throughout the paper, capital letters denote random variables,
and lower case letters denote the potential values they may take.
We do not explicitly state various technical measurability assumptions as these
can be deduced from the context.\footnote{
For simplicity, we could assume that $d$ takes on a countable set of values $\mathcal{D}$ or make
separability assumptions which imply that the stochastic process $\{Y_d, d \in \mathcal{D}\}$ is  defined
from its definition over a countable subset $\mathcal{D}_0 \subset \mathcal{D}$. See \citen{vdV-W}.}

The objective of causal or structural analysis is to learn about
features of the distributions of potential outcomes $Y_d$. Of
primary interest to us are the $\tau$-th quantiles of potential
outcomes under various treatments $d$, conditional on observed
characteristics $X=x$, denoted as $$ q(d, x, \tau).$$ We will
refer to the function $q(d,x,\tau)$ as the quantile treatment
response (QTR) function.  We are also interested in the quantile
treatment effects (QTE), defined as
$$
 q(d_1, x, \tau) - q(d_0, x, \tau),$$
that summarize the differences in the impact of treatments on the
quantiles of potential outcomes (\citen{lehmann:1974}, \citen{doksum:1974}).

Typically, the realized treatment $D$ is selected in relation to
potential outcomes, inducing endogeneity. This endogeneity
makes the conventional quantile regression of observed $Y$ on
observed $D$, which relies upon the restriction
$$
P[ Y \leq \theta(D, X, \tau) | X, D] = \tau \ \text { a.s., }
$$
inappropriate for measuring $q(d,x,\tau)$ and the
QTE.  Indeed the function  $\theta(d, x, \tau)$ solving these
equations will not be equal to  $q(d,x,\tau)$ under endogeneity.
The model presented next states conditions under
which we can identify and estimate the quantiles of latent
outcomes
 through the use of instruments $Z$ that affect $D$ but are independent of potential outcomes
 and the nonlinear quantile-type conditional moment
restrictions
$$
P[ Y \leq q(D, X, \tau) | X, Z] = \tau \ \text { a.s. }
$$

\subsection{The Instrumental Quantile Treatment Effects (IVQT) Model.}

Having conditioned on the observed characteristics $X=x$, each
latent outcome $Y_d$  can be related to its quantile function
$q(d,x,\tau)$ as\footnote{This follows by Fisher-Skorohod
representation of random variables which states that given a
collection of variables $\{\zeta_d\}$, each variable $\zeta_d$
can be represented as $ \zeta_d = q(d, U_d)$, for some $U_d\sim
U(0,1)$, cf. \citen{durrett}, where $q(d, \tau)$ denotes the
$\tau$-quantile of variable $\zeta_d$.}
\bsnumber\label{representation}
 Y_d =  q(d,x, U_d), \text{ where } U_d \sim U(0,1)
\end{split}\end{align}
is the structural error term.  We note that
representation (\ref{representation}) is essential to what
follows.

The structural error $U_d$ is responsible for heterogeneity of potential
outcomes among individuals with the same observed characteristics
$x$. This error term determines the relative
ranking of observationally equivalent individuals in the distribution of potential outcomes
given the individuals' observed characteristics, and thus we refer to $U_d$ as the rank variable.
Since $U_d$ drives differences in observationally equivalent individuals, one may think of $U_d$
as representing some unobserved characteristic, e.g.
ability or proneness.\footnote{\citen{doksum:1974} uses the term
proneness as in ``prone to learn fast" or ``prone to grow
taller".} This interpretation makes quantile analysis an
interesting tool for describing and learning the structure of
heterogeneous treatment effects and accounting for unobserved
heterogeneity; see \citen{doksum:1974}, \citen{heckman:smith}, and \citen{koenker:book}.

For example, consider a returns-to-training model, where $Y_d$'s
are potential earnings under different training levels $d$, and
$q(d,x,\tau)$ is the conditional earnings function which describes how an
individual having training $d$, characteristics $x$, and the latent ``ability" $\tau$ is rewarded
by the labor market. The earnings function may be different for
different levels of $\tau$, implying heterogeneous effects of
training on earnings of people that have different levels of
``ability". For example, it may be that the largest returns to training accrue to
those in the upper tail of the conditional distribution, that is, to the
``high-ability" workers.\footnote{It is important to note that the quantile index, $\tau$, in $q(d,x,\tau)$ refers to the quantile of potential outcome $Y_d$ given that exogenous variables are set at $X = x$ and not to the unconditional quantile of $Y_d$.  For example, suppose that one of the control variables in the earnings example is years of schooling.  An individual at the 30$^{\textnormal{th}}$ percentile of the distribution of $Y_d$ given say 20 years of schooling is not necessarily low income as even a relatively low earner with that level of education may still earn above the median earnings in the overall population.}

Formally, the IVQT model consists of five conditions (some are
representations) that hold jointly.

\noindent \textbf{Main Conditions of the Model:} Consider a common probability space $(\Omega, F, P)$
and  the set
of potential outcome variables $(Y_d, d \in \mathcal{D})$,  the
covariate variables $X$, and the instrumental variables $Z$.  The following
conditions  hold \textit{jointly} with probability one:
\begin{itemize}
\item[\textbf{A1 }]\textsc{Potential Outcomes.}  Conditional on $X$ and
for each $d$, $Y_{d} = q(d, X, U_{d})$, where $\tau \mapsto q(d, X, \tau)$ is non-decreasing on $[0,1]$
and left-continuous and $U_d \ {\sim} \ U(0,1)$.
\vspace{.1in}
\item[\textbf{A2 }]\textsc{Independence.  } Conditional on $X$ and for each $d$,
  $U_{d}$  is independent of instrumental variables $Z$.
\vspace{.1in}
\item[\textbf{A3 }]\textsc{Selection.} $D :=
\delta (Z, X, V)$ for some unknown function $\delta$ and random
vector $V$. \vspace{.1in}
\item[\textbf{A4 }]\textsc{Rank Similarity.}
Conditional on
$(X, Z, V)$,  $\{U_d\} \text{ are identically distributed. }$  \vspace{.1in}
\item[\textbf{A5 }] \textsc{Observed} random vector consists of \
$Y := Y_D, \  D , \ X$ and $Z.$
\end{itemize}

The following is the main econometric implication of the model.

\begin{theorem}[Main Statistical Implication] Suppose conditions
A1-A5 hold. (i) Then we have for $U:= U_D$, with probability one,
\begin{equation} \label{main implication}
Y = q(D,X,U), \ \   U \sim U(0,1) |X,Z.
\end{equation}
(ii) If (\ref{main implication}) holds and $\tau \mapsto q(d,\tau)$ is strictly increasing for each $d$, then for each $\tau \in (0,1)$, a.s
 \bsnumber\label{e1}
P\[Y \leq q(D, X, \tau)|X, Z \]  = \tau.
  \end{split}\end{align}
(iii)  If (\ref{main implication}) holds, then for any closed subset $I$ of $[0,1]$, a.s.
 \bsnumber\label{e1b}
P (U \in I) \leq  P\[ Y \in  q(D,X,I)|X, Z \],
  \end{split}\end{align}
  where $q(d,x,I)$ is the image of $I$ under the mapping $\tau \mapsto q(d,x,\tau)$.
\end{theorem}

The first result states that the main consequence of A1-A5 is a simultaneous equation model (\ref{main implication})
with non-separable error $U$ that is independent of $Z,X$, and normalized so that $U \sim U(0,1)$.  The second result considers
econometric implications when $\tau \mapsto q(D,X,\tau)$ is strictly increasing, which requires that $Y$ is non-atomic conditional on $X$ and $Z$.  In this case, we obtain the conditional moment restriction (\ref{e1}).
This implication follows from the first result and the fact that
$$
\{ Y \leq q(D,X,\tau) \} \text{ is equivalent to }  \{ U \leq
\tau\},
$$
when $q(D,X,\tau)$ is strictly increasing in $\tau$.  The final result deals with the case where  $Y$ may have atoms conditional on $X$ and $Z$, e.g. when $Y$ is a count or discrete response variable.  The first two results were obtained in  \citen{iqr:ema}, and the third result
is in the spirit of results given in \citen{ChesherRosenSmolinski}; \citen{ChesherDiscrete}; and \citen{ChesherSmolinski}. The latter results
are related to random set/optimal transport methods for identification analysis; see \citen{BMM:2011};  \citen{EGH:2010}; \citen{GH:2009}; and \citen{GH:2011}.

The model and the results of Theorem 1 are useful for two
reasons. First, Theorem 1 serves as a means of identifying  the
QTE in a reasonably general heterogeneous effects model. Second,
by demonstrating that the IVQT model leads to the conditional
moment restrictions (\ref{e1}) and (\ref{e1b}), Theorem 1 provides an economic and
causal foundation for estimation based on these restrictions.
%Thus, the IVQT model provides conditions under which one can
%recover the quantiles of potential outcomes from econometric
%equations (\ref{e1}) and (\ref{e1b}) in applications.

 \subsection{The Identification Regions.}
The conditions presented above yield the following identification region
for the structural quantile function $(d,x,\tau) \mapsto q(d,x,
\tau)$. The identification region for the case of strictly increasing
$\tau \mapsto q(d,x,
\tau)$  can be stated as the set $\mathcal{Q}$ of functions
$(d,x,\tau) \mapsto m(d,x,\tau)$ that satisfy the following relations, for all $\tau \in (0,1]$
\begin{equation}\label{e2}
  P[Y < m(D, X,\tau)|X, Z ] = \tau \text{ a.s. }
\end{equation}
This representation of the identification region $\mathcal{Q}$ is
implicit.  Nevertheless, statistical
inference about  $q \in \mathcal{Q}$ can be based on (\ref{e2}) and can
be carried out in practice using weak-identification robust
inference as described in
\citen{iqr:joe2}, \citen{SakataMarmer}, \citen{JunWeakIdIVQR}, \citen{SantosPartialIDIVQR}, or \citen{fsqr}.  Under conditions that yield point
identification, these regions collapse to a singleton, and the aforementioned weak-identification-robust inference
procedures retain their validity.

The identification region for the case of weakly increasing
$\tau \mapsto q(d,x, \tau)$ can be stated as the set $\mathcal{Q}$ of functions
$(d,x,\tau) \mapsto m(d,x,u)$ that satisfy the following relations:  For any closed subset $I$ of $(0,1]$,
 \bsnumber\label{e2b}
P (U \in I) \leq  P\[ Y \in m(D,X,I) |X, Z \]  \text{ a.s.},
  \end{split}\end{align}
where $m(D,X,I)$ is the image of $I$ under the mapping $\tau \mapsto m(D,X,\tau)$.   The inference problem here falls in the class of conditional moment inequalities and
approaches such as those described in \citen{AndrewsShi} or \citen{CLRIntersectionBounds}, for example, can be used.  The sets
$I$ to be checked could be reduced by determining approximate core-determining subsets;
see \citen{ChesherRosenSmolinski},  \citen{GH:2009}, \citen{GH:2011}  for further discussion.

\subsection{Discussion of the Model} Condition A1 imposes monotonicity on the structural
function of interest which makes its relation to the QTR apparent.
Condition A2 states that potential outcomes are independent of
$Z$, given $X$, which is a conventional independence restriction.
Condition A3 is a convenient representation of a treatment
selection mechanism, stated for the purposes of discussion. In A3,
the unobserved random vector $V$ is responsible for the
difference in treatment choices $D$ across observationally
identical individuals.  Dependendence between $V$ and $\{U_d\}$
is the source of endogeneity that makes the conventional exogeneity
assumption $U \sim U(0,1)|X,D$ break down.  This failure leads to inconsistency of
exogenous quantile methods for estimating the structural quantile function.  Within the model outlined above, this breakdown is resolved through the use of instrumental variables.

The independence imposed in A2 and A3 is
weaker than the commonly made assumption that
both the disturbances $\{U_d\}$ in the outcome equation and the
disturbances $V$ in the selection equation are \textit{jointly}
independent of the instrument $Z$; e.g. \citen{heckman:robb} and
\citen{late}. The latter assumption may be
 violated when the instrument is measured with error as discussed in
 \citen{hausman:1977} or the instrument is not assigned
 exogenously relative to the selection equation as in Example 2 in \citen{late}.

Condition A4 restricts the variation in ranks across potential outcomes and is key for identifying the QTR and associated QTE. Its
simplest, though  strongest, form is rank invariance, when
ranks $U_d$ do not vary with potential treatment states
$d$:\footnote{Notice that under rank invariance, condition A3 is a
pure representation, not a restriction, since nothing restricts
the unobserved information component $V$.}  \begin{equation} \label{invariance}
U_d =U \text{ for each } d \in \mathcal{D}.
\end{equation}
For example, under rank invariance, people who are strong (highly
ranked) earners without a training program ($d=0$) remain strong earners
having done the training ($d=1$). Indeed, the earnings of a person with characteristics $x$ and
rank $U=\tau$ in the training state ``0" is $Y_0 = q(0, x, \tau)$
and in the state ``1" is $Y_1 =q(1, x, \tau)$.\footnote{Rank
invariance is used in many interesting models without
endogeneity. See e.g. \citen{doksum:1974}, \citen{heckman:smith}, and
\citen{koenker:geling}.} Thus, rank invariance implies that a
common unobserved factor $U$, such as innate ability, determines
the ranking of a given person across treatment states.

Rank invariance implies that the potential outcomes $\{
Y_d\}$ are jointly degenerate
which may be implausible on logical grounds, as pointed out by \citen{heckman:smith}. Also, the rank variables $U_d$ may be
determined by many unobserved factors. Thus,  it is desirable to
allow the rank $U_d$ to change across $d$, reflecting some
unobserved, asystematic variation. Rank similarity A4
achieves this property while managing to
preserve the useful moment restriction (\ref{e1}).

Rank similarity A4 relaxes exact rank invariance by allowing
asystematic deviations, ``slippages" in the terminology of \citen{heckman:smith},  in one's rank away from
some common level $U$. Conditional on $U$, which may enter
disturbance $V$ in the selection equation, we have the following condition on the
slippages\footnote{Conditioning is required to be on all
components of $V$ in the selection equation A3. } \be  U_d - U \ \
\text{are identically distributed across } d \in \mathcal{D}. \ee In
this
  formulation, we implicitly assume that one selects
  the treatment  without knowing the exact potential outcomes; i.e. one
  may know $U$ and even the distribution of slippages, but does not know the
exact
   slippages $U_d-U$.  This assumption is consistent with many empirical
   situations where the exact latent outcomes are not
   known before receipt of treatment.  We also note that conditioning on appropriate covariates $X$
 may be important to achieve rank similarity.

 In summary, rank similarity is an important restriction of the IVQT model that
allows us to address endogeneity. This restriction is absent in
conventional endogenous heterogeneous treatment effect models.
However, similarity enables a more general selection mechanism,
A3, and weaker independence conditions on instruments than often are assumed in nonseparable IV models. The main
force of rank similarity and the other stated assumptions is the
implied moment restriction (\ref{e1}) of Theorem 1, which is
useful for identification and estimation of the quantile
treatment effects.

\subsection{Examples}  We present some examples that highlight the nature of the model,
its strengths, and its limitations.

\begin{example}[Demand with Non-Separable Error] The
following is a generalization of the
classic supply-demand example. Consider the model
 \bsnumber\label{demand}
\begin{array}{llll}
 & Y_{p} = q\(p, U\), \\
 & \tilde Y_{p} = \rho\(p, z, \mathcal{U}\), \\
  & P \ \in \{p:  \rho\(p, Z, U\) =  q\(p, \mathcal{U}\) \},
\end{array}
\end{split}\end{align}
where functions $q$ and $\rho$ are increasing in the last
argument. The function $p \mapsto Y_p$ is the random demand
function, and $p \mapsto \tilde Y_p$ is the random supply
function. Additionally, functions $q$ and $\rho$ may depend on
 covariates $X$, but this dependence is suppressed.

 Random variable $U$ is the level of demand and describes the
 demand curve at different states of the world. Demand is maximal
 when $U=1$ and minimal when $U=0$, holding $p$
fixed. Note that we imposed rank invariance (\ref{invariance}), as is typical in classic supply-demand models, by
making $U$ invariant to $p$.

Model (\ref{demand}) incorporates traditional additive
error models for demand which have $Y_p = q(p) + \epsilon$ where $\epsilon = Q_{
\mathcal{\epsilon}}(U)$. The model is much more general in that
the price can affect the entire distribution of the demand curve,
while in traditional models it only affects the location of the
distribution of the demand curve.

The $\tau$-quantile of the demand curve $p \mapsto Y_p$ is given
by $ p \mapsto  q(p, \tau ). $ Thus, the
curve $p \mapsto Y_p$ lies below the curve $p \mapsto  q(p, \tau
)$  with probability $\tau$.  Therefore, the various quantiles of the potential outcomes
play an important role in describing the distribution and
heterogeneity of the stochastic demand curve. The quantile
treatment effect may be characterized  by $
\partial q(p, \tau )/ \partial p
$ or by an elasticity $
\partial \ln q(p, \tau )/ \partial \ln p.
$ For example, consider the Cobb-Douglas model $ q(p, \tau) = \exp
\( \beta(\tau) + \alpha(\tau) \ln p     \)$ which corresponds to
a Cobb-Douglas model for demand with non-separable error $ Y_p =
\exp ( \beta(U) + \alpha( U)\ln p ). $ The log transformation
gives $ \ln Y_p = \beta(U) + \alpha( U) \ln p, $ and the quantile
treatment effect for the log-demand equation is given by the
elasticity of the original $\tau$-demand curve $ \alpha(\tau) =
\frac{\partial Q_{\sss \ln Y_p}(\tau) }{\partial \ln p} =
\frac{\partial \ln q(p, \tau) }{\partial \ln p}. $

The elasticity $\alpha(U)$ is random and depends on the state of
the demand $U$ and may vary considerably with $U$. For example,
this variation could arise when the number of buyers varies and
aggregation induces a non-constant elasticity across the demand
levels. \citen{iqr:joe2} estimate a simple demand model based on data from a New York fish
market that was first collected and used by
\citen{graddy}.  They find point estimates of the demand elasticity, $\alpha(\tau)$,
that vary quite substantially from $-2$ for low quantiles to $-0.5$
for high quantiles of the demand curve.

The third condition in (3.3), $P \ \in \{p:  \rho\(p, Z, U\) =
q\(p, \mathcal{U}\) \}$, is the equilibrium condition that
generates endogeneity; the  selection of the clearing price $P$
by the market depends on the potential demand and supply
outcomes. As a result  we have a representation that is
consistent with A3,
$ P = \delta ( Z, V), $ where $V$ consists of $U$ and $\mathcal{U}$ and may include "sunspot" variables if the equilibrium price is not unique. Thus what we observe can be
written as
 \bsnumber\label{simultaneous}
& Y  := q (P, U),  \  \  P := \delta ( Z, V), \ \ U
\text{ is independent of } Z.
\end{split}\end{align}

Identification of the $\tau$-quantile of the demand function, $p
\mapsto q(p, \tau)$ is obtained through the use of instrumental
variables $Z$, like weather conditions or factor prices, that
shift the supply curve and do not affect the level of the demand
curve, $U$, so that independence assumption A2 is met.
Furthermore, the IVQT model allows arbitrary correlation between
$Z$ and $V$. This property is important as it allows, for example,
$Z$ to be measured with error or to be exogenous relative to the
demand equation but endogeneous relative to the supply equation.

\end{example}

\begin{example}[Savings] \citen{401k} use the framework of the IVQT model to examine
 the effects of participating in a 401(k) plan on an
individual's accumulated wealth.  Since wealth is continuous,
wealth, $Y_d$, in the participation state $d \in \{0,1\}$ can be
represented as
$$
Y_d = q(d, X, U_d),  \ \ U_d \sim U(0,1)
$$
where $\tau \mapsto q(d, X, \tau)$ is the conditional quantile function of
$Y_d$ and $U_d$ is an unobserved random variable.
$U_d$ is an unobservable that drives differences in accumulated wealth conditional on $X$ under participation state $d$.  Thus, one might think of $U_d$ as the preference for saving and interpret the quantile
index $\tau$ as indexing rank in the preference for saving distribution.
One could also model the individual as selecting the 401(k) participation state to maximize expected utility:
 \bsnumber\label{roy}
D & =  \arg \max_{d \in \mathcal{D}} \  E\[ \ W \{ Y_d, d \} \Big
| X,Z,V
\]
   =  \arg \max_{d \in  \mathcal{D}}  \ E\[ \ W \{q (d,
x,  U_d), d\}  \Big | X,Z,V \], \\
\end{split}\end{align}
where $W\{Y_d,d\}$ is the random indirect utility derived under participation state $d$.\footnote{It may
depend on both observables in $X$ as well as realized and unrealized unobservables. Only
dependence on $Y_d$ and $d$ is highlighted.} As a result, the participation decision is represented by
$$ D = \delta (Z, X, V), $$
where $Z$ and $X$ are observed, $V$ is an unobserved information
component that may be related to ranks $U_d$ and includes other unobserved
variables that affect the participation state, and function $\delta$ is
unknown. This model fits into the IVQT model with the independence condition A2
requiring that $U_d$ is independent of $Z$, conditional on $X$.

The simplest form of rank similarity is rank invariance (\ref{invariance}), under which the preference for
saving vector ${U_d}$ may be collapsed to a single random variable $U = U_0 = U_1.$
In this case, a single preference for saving is responsible for an individual's ranking
across all treatment states. The rank similarity condition A4 is a more general form of rank
invariance.  It relaxes the exact invariance of ranks $U_d$
across $d$ by allowing noisy, unsystematic variations of $U_d$
across $d$, conditional on $(V,X,Z)$.
This relaxation allows for variation in rank across the
treatment states, requiring only an ``expectational rank invariance."  Similarity implies that
given the information in $(V,X,Z)$ employed to make the selection of treatment
$D$, the expectation of any function of rank $U_d$ does not vary across the
treatment states. That is, ex-ante, conditional on $(V,X,Z)$, the ranks may be
considered to be the same across potential treatments, but the realized, ex-post,
rank may be different across treatment states.

From an econometric perspective, the similarity assumption is
nothing but a restriction on the unobserved
heterogeneity component which precludes systematic variation of
$U_d$ across the treatment states. To be more concrete, consider the following
simple example where $$ U_d = F_{V+\eta_d}(V +
\eta_d),
$$
where $F_{V+\eta_d}(\cdot)$ is the distribution function of $V+\eta_d$ and
$\{\eta_d\}$ are mutually iid conditional on $V$, $X$,
and $Z$. The
variable $V$ represents an individual's ``mean" saving preference,
while $\eta_d$ is a noisy adjustment.\footnote{Clearly similarity
holds in this case, $U_d \overset{d}= U_{d'}$ given $V$, $X$, and $Z$.}
This more general assumption leaves the individual optimization problem (\ref{roy})
unaffected, while allowing variation in an individual's rank
across different potential outcomes.

While we feel that similarity may be a reasonable assumption in many contexts,
imposing similarity is not innocuous.  In the context of 401(k)
participation, matching practices of employers could
jeopardize the validity of the similarity assumption.    To be more concrete, let $U_d = F_{V+\eta_d}(V +
\eta_d)$ as before but let $\eta_d = d M$ for random variable $M$ that depends on the match rate and is independent
of $V$, $X$, and $Z$.  Then conditional on $V = v$, $X$, and $Z$, $U_0 = F_V(v)$ is degenerate but $U_1 = F_{V+M}(v+M)$ is not.  Therefore, $U_1$ is not equal to $U_0$ in distribution.
Similarity may still
hold in the presence of the employer match if the rank, $U_d$, in the asset distribution
is insensitive to the match rate.  The rank may be insensitive if, for example, individuals follow
simple rules of thumb such as target saving when they make their savings decisions.  Also,
if the variation of match rates is small relative to the variation of individual heterogeneity
or if the covariates capture most of the variation in match rates,
then similarity may be satisfied approximately.

\end{example}

\begin{example}[Discrete Choice Model with Market-Level Data] \citen{BerryHaileDiscreteChoiceId} show that a general model for market-level data realized from a discrete-choice problem can fit within the IVQT model.  To keep notation and exposition simple, we consider a much-simplified version of the model from \citen{BerryHaileDiscreteChoiceId} in which consumer $i$'s indirect utility from choosing product $j$ is
$$U_{ijt} = u(X_{jt},P_{jt},\xi_{jt},V_{ijt}) = u(\delta_j(X_{jt},\xi_{jt}),P_{jt},V_{ijt}),$$
where $t$ indexes markets, $X_{jt}$ are observed exogenous product-market characteristics, $P_{jt}$ is the observed price of product $j$ in market $t$ which is treated as endogenous, $\xi_{jt}$ are product-market specific unobservables, and $V_{ijt}$ are individual-product-market specific unobservables that have density $f(\cdot)$.  Thus, the model imposes that unobserved product-market specific effects and observed variables $X_{jt}$ may only affect utility through the index $\delta_{jt} = \delta_j(X_{jt},\xi_{jt})$, where $\delta_j(\cdot, \cdot)$ may differ arbitrarily across products but is the same across all markets.  That unobserved product characteristics affect utility only through a scalar index is a substantive restriction but is common in the literature on discrete choice models where, for example, one can interpret the index as an aggregate representing product quality.

An individual will then choose the product that maximizes individual utility.  Letting $Y_{it}$ denote the observed choice of individual $i$, we have that
$$
Y_{it} = \arg \max_{j \le J} U_{ijt},
$$
where we assume the same $J$ products are available in each market for simplicity.\footnote{Obviously, identification of the model requires normalizations.  For example, the utility from one of the options is generally normalized to zero.  As this model is not the focus of this review, we do not discuss these normalizations which are discussed in detail in a more general context in \citen{BerryHaileDiscreteChoiceId}.}
The market share of each product will then be given as
\begin{align*}
S_{jt} &= \int 1\{u(\delta_{jt},P_{jt},v)= \max_{k\leq J} u(\delta_{kt},P_{kt},v) \}f(v)dv \\
&:= s_j(\{\delta_{jt},P_{jt}\}_{j = 1}^{J}) = s_j(\delta_t,P_t),
\end{align*}
where $\delta_t = (\delta_{1t},...,\delta_{Jt})'$ and $P_t = (P_{1t},...,P_{Jt})'$.

To fit this model into the instrumental variables quantile regression model, \citen{BerryHaileDiscreteChoiceId} make several assumptions to produce a structural relationship which is monotonic in a scalar unobservable.  First, they assume that the utility function $u(\delta_{jt},P_{jt},V_{ijt})$ is strictly increasing in $\delta_{jt}$.  This assumption is standard in the discrete choice literature and coincides with the interpretation of $\delta_{jt}$ as product quality where higher quality products are associated with higher utility all else equal.  Monotonicity of the utility function is not sufficient due to the fact that all that is observed is the market share which depends on the utility of each potential choice. Thus, \citen{BerryHaileDiscreteChoiceId} make an additional assumption that they term ``connected substitutes.''  Intuitively, this condition implies that an increase in the quality of every good within some strict subset of the available choices will be associated with the total market share of all goods not in the subset decreasing as long as the quality of no good outside of the subset increases.  \citen{BerryHaileDiscreteChoiceId} show that the connected substitutes condition is satisfied in usual random utility discrete choice models and that it can hold fairly generally.   Using these assumptions, \citen{BerryHaileDiscreteChoiceId} use a result from \citen{Gandhi2008} which shows that the system of equations
$$
S_{jt} = s_j(\delta_t,P_t)
$$
has a unique solution for the vector $\delta_t$ as long as all goods present in equilibrium have positive market shares.  Thus, we may write
\begin{align}
\label{InverseShare}
\delta_{jt} = g_j(S_t,P_t)
\end{align}
for some function $g_j$ where $S_t = (S_{1t},...,S_{Jt})'$.

From (\ref{InverseShare}), we have that $\delta_j(X_{jt},\xi_{jt}) = g_j(S_t,P_t)$.  To complete the argument, \citen{BerryHaileDiscreteChoiceId} assume that the function $\delta_j(X_{jt},\xi_{jt})$ is strictly increasing in its second argument, $\xi_{jt}$, which represents unobserved product attributes.  This condition rules out the case where $\xi_{jt}$ can represent attributes that would increase utility for some individuals but decrease utility for others and again corresponds to the notion that $\xi_{jt}$ represents unobserved product quality in which an increase unambiguously makes the product more desirable.  With the assumed monotonicity in the function $\delta_j$, one obtains
\begin{align*}
\xi_{jt} = \delta_j^{-1}(g_j(S_t,P_t);X_{jt}) = h_j(S_t,P_t,X_{jt}).
\end{align*}
It is also clear that $h_j(X_t,P_t,S_t)$ is strictly increasing in $S_{jt}$, which is proven in Lemma 5 of \citen{BerryHaileDiscreteChoiceId}, from which it follows that
\begin{align*}
S_{jt} = q_j(S_{-jt},P_t,X_{jt},\xi_{jt}),
\end{align*}
where $S_{-jt}$ denotes the set of market shares for each product in market $t$ excluding product $j$ and $q_j$ is an unknown function that is strictly increasing in $\xi_{jt}$.  Then, $q_j$ can be taken as the structural function in the instrumental variables quantile model after the normalization that $\xi_{jt}$ follows a $U(0,1)$, assuming that $\xi_{jt}$ has an atomless distribution.  The model is then completed by assuming the existence of instruments, $Z_t$, that are independent of $\xi_{jt}$ conditional on $X_{jt}$ and are related to the endogenous variables through $(S_{-jt}',P_t')' = \Delta(Z_t,X_{jt},V_t)$ for some function $\Delta$ and unobservables $V_t$.  Finally, note that the model assumes rank invariance in its construction.
\end{example}

\section{The Identifying Power of IV Quantile Restrictions}

The purpose of this section is to examine the identifying power of conditional
moment restrictions (\ref{e1}). Specifically, we give
various  conditions for point identification in this section, summarizing
and updating some of the results known in the literature.   We remark here that point
identification is not required in applications in principle as there exist inference methods
that apply without point identification. However, it is useful to know and understand conditions
under which moment conditions are informative enough that the identification
region shrinks to a single point; in such cases the inference methods will also
produce very informative confidence sets.  We present point-identifying conditions first for the binary case, $D \in \{0,1\}$
and $Z \in \{0,1\}$, and then consider the case
of $D$ taking a finite number of values, and finally consider the continuous case.

\subsection{Conditions for point identification in the binary case}  Here we consider
the cases where $D \in \{0,1\}$
and $Z \in \{0,1\}$. The following analysis is all conditional on $X=x$ and for a given
quantile $\tau \in (0,1)$, but we suppress this dependence for
ease of notation. Under the conditions of Theorem 1, we know that
there is at least one function $q(d):= q(d, x, \tau)$ that
solves  $ P[Y \leq q(D)|Z] = \tau \ \ \text{a.s}.$ The function
$q(\cdot)$ can be equivalently represented by a vector of its
values $q=(q(0), q(1))'$. Therefore, for vectors of the form $y =
(y_0, y_1)'$, we have a vector of moment equations
 \begin{equation}\label{2eq}
\Pi (y) := \big ( \ P[Y \leq y_D|Z=0] - \tau, \ P[Y \leq
y_D|Z=1] - \tau  \ )'
\end{equation}
where $y_D := (1-D)\cdot y_0 + D \cdot y_1$.
We say that $q$ is identified in some parameter space, $\mathcal{L}$, if
$y=q$ is the only solution to $\Pi(y)=0$ among all $y \in \mathcal{L}$.

We require that the Jacobian
$\partial \Pi(y)$ of $\Pi(y)$ with respect to $y=(y_0, y_1)'$ exists and that it takes the form
\begin{eqnarray} \partial \Pi(y) & := &  \left[
\begin{array}{cc}
 f_{\sss Y}(y_0|D=0,Z=0) P[D=0|Z=0]  &  f_{\sss Y}(y_1|D=1,Z=0) P[D=1|Z=0]  \\
   f_{\sss Y}(y_0|D=0,Z=1) P[D=0|Z=1]  &  f_{\sss Y}(y_1|D=1,Z=1) P[D=1|Z=1]
\end{array} \right ] \nonumber \\
& =: &  \left[
\begin{array}{cc}
 f_{\sss Y,D}(y_0, 0|Z=0) &  f_{\sss Y,D}(y_1, 1| Z=0)   \\
   f_{\sss Y, D}(y_0, 0|Z=1) &  f_{\sss Y,D}(y_1,1|Z=1) \label{eq:defJ}
\end{array} \right ].
\end{eqnarray}

For local identification, we take $\mathcal{L}$ as an open neighborhood of $q = (q(0), q(1))'$.
For global identification, we shall use some definitions from Mas-Collell to define
$\mathcal{L}$. In what follows, for every proper (non-null) subspace $L \subset \mathbb{R}^l$,
 let $\textrm{proj}_L : \mathbb{R}^l \mapsto L$ denote the perpendicular projection map. A convex, compact
 polytope is  a bounded convex set formed by an intersection of a finite number of closed half-spaces.
Such a polytope is of full dimension in $\mathbb{R}^l$ if it has a non-empty interior in $\mathbb{R}^l$.
  A face of a polytope $\mathcal{L}$ is the intersection of any supporting hyperplane
   of $\mathcal{L}$ with $\mathcal{L}$, so that
  faces of a polytope necessarily include the polytope itself. For instance, a rectangle in
  $\mathbb{R}^2$ has one 2-dimensional face given by itself, four 1-dimensional faces given by its edges, and
  four 0-dimensional faces gives by its vertices. A subspace spanned by a non-empty face of $\mathcal{L}$
  is the translation to the origin of the minimal affine space containing that face.

\begin{theorem} [Identification by Full Rank Conditions]\label{theorem: binary id}
Suppose that $\Pi(q) =0$, the support of $D$ is $\{0,1\}$ and the
support of $Z$ is $\{0,1\}$.  Assume  that the conditional density $f_{Y}( y |D=d,Z=z)$ exists for each $y \in \mathbb{R}$
and $(d,z) \in \{0,1\} \times \{0,1\}$. (i) (Local) Suppose  the Jacobian
 $\partial \Pi$ given by (\ref{eq:defJ}) is continuous and has full rank at $y=q$, then the $\tau$-quantiles of potential outcomes,
$q=(q(0), q(1))'$, are identified in the region $\mathcal{L}$ given by  a sufficiently small open neighborhood of $q$ in $\mathbb{R}^2$. (ii) (Global) Assume that region $\mathcal{L}$ contains $q$ and can be covered by a finite number of compact convex 2-dimensional polytopes $\{\mathcal{L}_j\}$, each containing $q$. Assume that for each $j$, $\partial \Pi$ is a $C^1$ Jacobian of $\Pi:  \mathcal{L}_j \to \mathbb{R}^2$ , and that, possibly after rearranging the rows of $\partial \Pi$,  for each $y \in \mathcal{L}_j$ and each subspace
$L \subset \mathbb{R}^2$ spanned by a face of $\mathcal{L}_j$ that includes $y$, the linear map $$\textrm{proj}_{L}  \circ \partial \Pi(y): L \mapsto L$$ has a positive determinant. Then $q$ is identified in $\mathcal{L}$.
\end{theorem}

The first result is a simple local identification condition of the type considered in \citen{rothenberg:1971} which we provide to fix ideas.
The second result  is a global identification condition which extends the result in \citen{iqr:ema} by allowing non-rectangular sets $\mathcal{L}$. This result is based on the global univalence theorems of \citen{mas-colell:convex}. As explained below, the positive determinant condition  requires the impact of instrument $Z$ on the joint distribution
of $(Y,D)$ to be sufficiently rich. In particular,  the
instrument $Z$ should not be independent of the endogenous
variable $D$. We note that  existence of the conditional density $f_{Y}( y |D=d,Z=z)$ is only required
for $(d,z)$ in the support of $(D,Z)$. Outside the support we can define the conditional density as 0, so the existence condition
is not very restrictive.   Moreover, the condition is formulated so that
$\mathcal{L}$ can take on relatively rich shapes that can carry  useful economic restrictions.   For instance,
in the training context, a useful restriction on the parameters is that training weakly increases the potential earning quantiles. This restriction  can be implemented by taking some natural parameter space and intersecting it with the half-space $H= \{ (y_0, y_1) \in \mathbb{R}^2: y_1 \geq y_0\}$. Specifically, a cube $C = \{ y \in \mathbb{R}^l:  \| y\|_{\infty} \leq K\}$ intersected with the halfspace
$H$ is an example of a region $\mathcal{L}$
permitted by the global identification result (ii).

\begin{remark}[Simple Sufficient Conditions] To illustrate the conditions of the theorem, let us consider
the parameter space $\mathcal{L}$ as either $\mathcal{L} = q+ C$, i.e. a cube centered at $q$,
or $ \mathcal{L} = (q + C) \cap H$, i.e. intersection of  a cube centered at $q$ with the halfspace $H$.  Consider the trivial covering
of $\mathcal{L}$ by itself, i.e. $\mathcal{L}_j = \mathcal{L}$.  Then the positive determinant condition of the theorem is implied by the following
simple conditions:
\begin{equation}\label{LMR}
\frac{ f_{\sss Y,D}(y_1,1|Z=1) }{f_{\sss Y, D}(y_0, 0|Z=1) } >
\frac {f_{\sss Y,D}(y_1,1|Z=0)}{  f_{\sss Y,D}(y_0,0|Z=0)} \text{
for all $y =(y_0, y_1)$ } \in \mathcal{L},
\end{equation}
and
\begin{equation}\label{POS}
 f_{\sss Y,D}(y_1,1|Z=1)  > 0,  \ \  f_{\sss Y,D}(y_0,0|Z=0) > 0, \ \ \text{
for all $y =(y_0, y_1)$ } \in \mathcal{L}. \end{equation}
Alternatively, since we can rearrange the rows of $\partial \Pi$, which corresponds to reordering elements of vector $\Pi$,
the positive determinant condition of the theorem is implied by the following simple conditions:
\begin{equation}\label{LMR2}
\frac{ f_{\sss Y,D}(y_1,1|Z=1) }{f_{\sss Y, D}(y_0, 0|Z=1) } <
\frac {f_{\sss Y,D}(y_1,1|Z=0)}{  f_{\sss Y,D}(y_0,0|Z=0)} \text{
for all $y =(y_0, y_1)$ } \in \mathcal{L},
\end{equation}
and
\begin{equation}\label{POS2}
f_{\sss Y,D}(y_1,1|Z=0)  > 0,  \ \  f_{\sss Y, D}(y_0, 0|Z=1) > 0, \ \ \text{
for all $y =(y_0, y_1)$ } \in \mathcal{L}. \end{equation}
The proof that these are sufficient conditions is given in the appendix, and below we discuss the economic plausibility
of these conditions.
\end{remark}

\begin{remark}[Plausibility of (\ref{LMR}) and (\ref{POS})]
The condition (\ref{POS}) seems quite mild, so we focus on (\ref{LMR}).  We can illustrate (\ref{LMR}) by considering the problem of evaluating a training program where $Y$'s are
earnings, $D$'s $\in \{0,1\}$ are training states, and $Z$'s $\in
\{0,1\}$ are offers of training service.  Condition (\ref{LMR}) may be interpreted as a \textit{monotone
likelihood ratio} condition. That is, the instrument $Z$ should
have a monotonic impact on the  likelihood ratio specified in
(\ref{LMR}).  This monotonicity may be a weak condition in some contexts
and a strong condition in others.  For instance, if $\mathcal{L}$
is a cube $q+ C$, then this condition
may be considered relatively strong. On the other hand, if we impose
monotonicity of the training impact on earning quantiles, so that
$q(0) \leq q(1)$, i.e. $q \in  \mathcal{L} =  (q+ C) \cap H$, then
condition (\ref{LMR}) would be trivially satisfied in many empirical settings. Indeed, it would
suffice that the instrument $Z$, the offer of training
services, increases the relative joint likelihood of receiving
higher earnings and receiving the training service.  In
many instances, we also have $P[D=1|Z=0]=0$; e.g. those not
offered training services do not receive that training.  When $P[D=1|Z=0]=0$, the right-hand side of (\ref{LMR}) equals $0$ which makes the
identification condition (\ref{LMR}) satisfied trivially even for
the less convenient parameter sets such as $\mathcal{L} = q+ C$. \end{remark}

\subsection{Identification with Multiple Points of Support}
We generalize the result of Theorem 2 to
more general discrete treatments with discrete instruments.
Consider the case when $D$ has the support
$\{1,...,l\}$ and $Z$ has the support $\{1,...,r\}$ ($l \leq r <
\infty$). Note that function $q(\cdot)$ can be represented by a
vector $q=(q(1),...,q(l)) '\in \mathbb{R}^l$. Under the conditions of
Theorem 1, there is at least one function $q(d)$ that solves $ P[Y
\leq q(D)|Z] = \tau \ \ \text{a.s}.$ Therefore, for vectors of
the form $y = (y_1,...,y_l)'$ and the vector of moment equations
  \bsnumber\label{req}
\Pi (y) = \big ( P[Y \leq y_{D}|Z=z] - \tau, \ \ z =1,..., r)',
\end{split}\end{align}
where $y_{D} := \sum_{d} 1[D=d]\cdot y_d$, the model is identified if $y=q$ uniquely solves $\Pi (y)=0$.

We define matrix $\partial \Pi(y)$ as the $r \times l$
matrix with $(d,z)$ element given by  $f_{\sss Y}(y_d|D=d,Z=z)
P[D=d|Z=z]$ where $z=1,...,r$ and $d =1,...,l$.   We require
this to be the Jacobian matrix of the map $y \mapsto \Pi(y)$ and impose
full-rank-type conditions on submatrices of this Jacobian.
To this end, let $m$ denote any permutation of $l$ distinct
integers from $\{1,...,r\}$, called $l$-permutations, and $\mathcal{M}$ be a collection of all such permutations.
Let $\Pi_m : = (\Pi_j)_{j \in m}$, which maps $\mathbb{R}^l$ to $\mathbb{R}^l$, be a subvector of $\Pi$ formed by selecting $j$-th elements
of $\Pi$ according to their order in $m$.\footnote{Note that this formulation allows reordering elements of $\Pi$ which may be needed to achieve the required positive determinant condition as discussed in the binary case.} Let $\partial \Pi_m$ denote the corresponding $l \times l$ Jacobian matrix of $\Pi_m$. The following theorem generalizes Theorem 2.

\begin{theorem}[Identification for Discrete $D$]\label{theorem:discrete D} Suppose $\Pi(q)=0$, the support of $D$ is
$\{1,...,l\}$ and of $Z$ is  $\{1,...,r\}$.  Assume  that the conditional density $f_{Y}( y |D=d,Z=z)$ exists for each $y \in \mathbb{R}$,
and $(d,z) \in \{1,...,l\} \times \{1,...,r\}$. (i) (Local) Suppose  the Jacobian
 $\partial \Pi(y)$ defined above is continuous and has rank $l$ at $y=q$. Then the $\tau$-quantiles of potential outcomes,
$q$, are identified in the region $\mathcal{L}$ given by  a sufficiently small open neighborhood of $q$ in $\mathbb{R}^l$.
 (ii) (Global) Assume that region $\mathcal{L}$ contains $q$ and can be covered by a finite number of compact convex $l$-dimensional polytopes $\{\mathcal{L}_j\}$, each containing $q$ and having the following properties:  For each $j$
 there is an $l$-permutation $m(j) \in \mathcal{M}$, such that  $\partial \Pi_{m(j)}$ is the $C^1$ Jacobian of $\Pi_{m(j)}:  \mathcal{L}_j \to \mathbb{R}^l$,  and for each $y \in \mathcal{L}_j$ and each subspace
$L \subset \mathbb{R}^l$ spanned by a face of $\mathcal{L}_j$ that includes $y$, the linear map $$\textrm{proj}_{L}  \circ \partial \Pi_{m(j)} (y): L \mapsto L$$ has a positive determinant. Then $q$ is identified in $\mathcal{L}$.
\end{theorem}

We note that in the theorem  existence of the conditional density $f_{Y}( y |D=d,Z=z)$ is only required
for $(d,z)$ in the support of $(D,Z)$. This density can be defined to take on an arbitrary value for $(d,z)$ outside the support.  The first result is a simple local identification condition provided to fix ideas. The second result  is a global identification condition based on Global Univalence Theorem 1 of \citen{mas-colell:convex}.
This result complements a similar result given in \citen{iqr:ema}  based on Global Univalence Theorem 2 of \citen{mas-colell:convex}. The positive determinant condition  requires the impact of instrument $Z$ on the joint distribution of $(Y,D)$ to be sufficiently rich.

\begin{remark}[An Alternative Sufficient Condition] Here we recall an alternative sufficient condition from  \citen{iqr:ema}, which is based
  on the Global Univalence Theorem 2 of \citen{mas-colell:convex}.  Assume that region $\mathcal{L}$ contains $q$ and can be covered by a finite number of compact convex $l$-dimensional sets $\{\mathcal{L}_j\}$, each containing $q$ and having the following properties: (i) For each $j$, there is a permutation $m(j) \in \mathcal{M}$ such that  $\partial \Pi_{m(j)}$  is $C^1$ Jacobian of $\Pi_{m(j)}:  \mathcal{L}_j \to \mathbb{R}^l$; (ii)
for each $y \in \mathcal{L}_j$,
 $$
 \det[\partial \Pi_{m(j)}(y)] >0;
 $$
(iii) $\mathcal{L}_j$ possesses a $C^1$-smooth boundary $\partial \mathcal{L}_j$;  and (iv) for each $y \in \partial \mathcal{L}_j$,  $l'(\partial \Pi_{m(j)}(y)+ \partial \Pi_{m(j)}(y)')l>0$ for each $l \in T (y): l\neq 0$ where  $T(y)$ is the subspace tangent to $\mathcal{L}_j$ at point $y$. Then $q$ is identified in $\mathcal{L}$.  This condition seems to require slightly stronger conditions on the boundary than the condition used in Theorem 3. The advantage of the conditions from \citen{iqr:ema} is that they more transparently convey the full-rank nature of the conditions imposed.
\end{remark}

\subsection{Identification with general D} Finally we consider
conditions for point identification in the case of more general $D$ and $Z$ that may take on a continuum of values.    We let
$d$ denote elements
in the support of $D$ and $z$ denote elements in the support of
$Z$.  Without loss of much
generality, we restrict attention to the case
where both $Y$ and $D$ have bounded support.  We require
the parameter space $\mathcal{L}$ to be a collection of bounded (measurable) functions
$m: \mathbb{R}^k \mapsto \mathbb{R}$ containing $q(\cdot)$.  We say that $q(\cdot)$ such that
$P[Y \leq q(D)|Z]= \tau$ a.s. is identified in $\mathcal{L}$ if  for any other $m(\cdot) \in
\mathcal{L}$ such that  $P[Y \leq m(D)|Z]=\tau$ a.s.,  $m(D)=q(D)$ a.s.  Below, we use $\|\cdot \|_{p,P}$
to denote the  $L^p(P)$ norm.

\begin{theorem}[Identification with General $D$]\label{theorem: general id} Suppose that $P[Y \leq q(D)|Z]= \tau$ a.s. and both $Y$ and $D$ have bounded
support. Consider a parameter space $\mathcal{L}$ which is a collection of bounded (measurable) functions
$m: \mathbb{R}^k \mapsto \mathbb{R}$ containing $q(\cdot)$. Assume that for
 $\epsilon:= Y- q(D)$ the conditional density $f_{\epsilon}( e |D,Z)$ exists for each $e \in \mathbb{R}$, a.s.
(i) (Global) Suppose that for each $\Delta(d) := m(d) -q(d)$ with $m(\cdot) \in
\mathcal{L}$, $\omega_{\Delta}(D,Z) := \int_0^1
f_{\epsilon}(\delta \Delta(D) |D, Z) d \delta >0$ a.s. and
\begin{equation}\label{global1}E\[ \Delta(D) \cdot \omega_{\Delta}(D,Z) |Z\]=0 \text{ a.s. }  \Rightarrow
\Delta(D)=0 \text{ a.s.}
\end{equation}
Then $q(\cdot)$ is identified in $\mathcal{L}$.
(ii) (Local) Suppose that $\omega_0(D,Z):= f_{\epsilon}(0 |D, Z)> 0$ a.s. and for each $\Delta(d) := m(d) -q(d)$ with $m(\cdot) \in
\mathcal{L}$,
\begin{equation}\label{local1}
E\[ \Delta(D) \cdot \omega_{0}(D,Z) |Z\]=0 \text{ a.s. }  \Rightarrow
\Delta(D)=0 \text{ a.s.},
\end{equation}
and, for some $0 \leq \eta <1$ and $1 \leq p$,
\begin{equation}\label{local2}
 \| E\[ \Delta(D) \cdot \{ \omega_{\Delta}(D,Z)- \omega_0(D,Z)\} |Z\] \|_{p,P}
\leq \eta  \| E \[ \Delta(D) \cdot \omega_0(D,Z) |Z\] \|_{p,P}.
\end{equation}
Then $q(\cdot)$ is identified in $\mathcal{L}$.
\end{theorem}

 Condition (i),   mentioned in \citen{iqr:ema}, states a  non-linear bounded completeness condition for global identification. The condition (\ref{global1}) required is not primitive, but it highlights a useful link with the linear bounded completeness
condition: $E\[ \Delta(D)|Z\]=0 \text{ a.s. }  \Rightarrow
\Delta(D)=0 \text{ a.s.}$ used by \citen{newey:powell}.  The latter condition is
needed for identification in the mean IV model $E [Y - q(D)|Z ] = 0$ under the assumption of a bounded structural function $q$.
The latter condition is known to be quite weak, as shown in \citeasnoun{Hault},
and there are many primitive sufficient conditions that imply this condition.  \citen{andrews:genericity}  shows
that linear completeness is generic under some conditions. Although condition (\ref{global1}) is not primitive, it is not vacuous either since the previous theorems provide primitive conditions for its validity.  The local identification condition (ii), obtained by \citen{CCLN}, provides yet another sufficient condition for condition (i). The result (ii) replaces the nonlinear completeness condition (\ref{global1}) by the linear
completeness condition (\ref{local1}) which is easier to check. The result (ii) also implicitly requires that the set $\mathcal{L}$ is a sufficiently small neighborhood of $q$ and that functional deviations $m(\cdot)-q(\cdot)$ and the conditional density $f_{\epsilon}(\cdot|D,Z)$ are sufficiently smooth. This is explained in detail in \citen{CCLN} where further primitive smoothness and completeness conditions are also provided.

\section{Other Approaches to Quantile Models with Endogeneity}

There are, of course, other sets of modeling assumptions that one could employ to build a quantile model with endogeneity.  In this section, we briefly outline two other approaches that have been taken in the literature.  The first, due to \citen{abadie}, extends the local average treatment effect (LATE) framework of \citen{late} to quantile treatment effects.  The second, considered in \citen{imbens:newey} and \citen{SLeeTriangularIVQR}, uses a triangular structure to obtain identification.

\subsection{Local Quantile Treatment Effects with Binary Treatment and Instrument}

In fundamental work, \citen{abadie} develop an approach to estimating quantile treatment effects within
the LATE framework of \citen{late} in the case where both the instrument and treatment variable are
binary.  The use of the LATE framework makes this approach appealing as many applied researchers are
familiar with LATE and the conditions that allow identification and consistent estimation of this
quantity.  Importantly, the extension proceeds under exactly the same monotonicity requirement as needed for LATE.

Specifically, \citen{abadie} show that the QTE for a subpopulation is identified if
\begin{itemize}
\item[1.] (Independence) the instrument $Z$
is independent of the potential outcome errors, $\{U_d\}$, and the errors in the selection
equation, $V$;
\item[2.] (Monotonicity) $P(D_1 \ge D_0 | X) = 1$ where $D_1$ is the treatment state of
an individual when $Z = 1$ and $D_0$ is defined similarly,
holds;
\item[3.] and other standard conditions are met.
\end{itemize}
The subpopulation for whom the QTE is identified is the set of ``compliers,'' those individuals with $D_1 > D_0$.  In other words, the compliers are the set of individuals whose treatment is altered by switching the instrument from zero to one.  Monotonicity is key in this framework.  The monotonicity condition rules out ``defiers,'' individuals who would receive treatment in the absence of the intervention represented by the instrument but would not receive treatment if placed into the treatment group.  The effects for individuals who would always receive treatment or never receive treatment regardless of the value of the instrument are unidentified.

Looking at these conditions, we see that the model of \citen{abadie} replaces the monotonicity assumption (A1), the independence assumption (A2), and the similarity assumption (A4) with a different type of monotonicity and a stronger independence assumption and identifies a different quantity:  the QTE for compliers.  The LATE-style approach has not yet been extended beyond cases with a binary treatment and a single binary instrument while the instrumental variable quantile model of \citen{iqr:ema} applies to any endogenous variables and instruments. Note that neither set of conditions nests the other, and neither framework is more general than the other.  Thus, the frameworks are best viewed as complements, providing two sets of conditions that can be considered when thinking about a strategy for estimating heterogeneous treatment effects.

Of course, the two sets of conditions may be mutually compatible.  One such case
is discussed in \citen{401k}.  In this example, the pattern of results obtained from the two
estimators is quite similar, and the difference between the estimates appears small relative to
sampling variation.  Further exploration of these two approaches and their similarities and differences may be interesting
to consider.

\subsection{Instrumental Variables Quantile Regression in Triangular Systems}

Another compelling framework is based on assuming a triangular structure as in \citen{imbens:newey}. See also \citen{chesher}, \citen{koenker:ma}, and \citen{SLeeTriangularIVQR} for related models and results.  The triangular model takes the form of a triangular system of equations
\begin{align*}
Y &= g(D,\epsilon), \\
D &= h(Z,\eta),
\end{align*}
where $Y$ is the outcome, $D$ is a continuous scalar endogenous variable, $\epsilon$ is a vector of disturbances, $Z$ is a vector of instruments with a continuous component, $\eta$ is a scalar reduced form error, and we ignore other covariates for simplicity.  It is important to note that
the triangular system generally rules out simultaneous equations which typically have
that the reduced form relating $D$ to $Z$ depends on a vector of disturbances.  For example, in a supply and demand system, the reduced form for both price and quantity will generally depend on the unobservables from both the supply equation and the demand equation.
Outside of $\eta$ being a scalar, the key conditions that allow identification of quantile effects in the triangular system are
\begin{itemize}
\item[1.] (Monotonicity) The function $\eta \mapsto h(Z,\eta)$ is strictly increasing in $\eta$, and
\item[2.] (Independence) $D$ and $\epsilon$ are independent conditional on $V$ for some observable or estimable $V$.
\end{itemize}

The variable $V$ is thus the ``control function'' conditional on which changes in $D$ may be taken as causal.  \citen{imbens:newey} use $V = F_{D|Z}(d,z) = F_{\eta}(\eta)$, where $F_{\eta}(\cdot)$ represents the CDF of $\eta$, as the control function and show that this variable satisfies the independence condition under the additional condition that $(\epsilon,\eta)$ is independent of $Z$.  They show that one may use $D = h(Z,\eta)$ to identify $V$ under the assumed monotonicity of $h(Z,\eta)$ in $\eta$.  Using $V$ obtained in this first step, one may then construct the distribution of $Y|D,V$.  Then integrating over the distribution of $V$ and using iterated expectations, one has
\begin{align*}
\int F_{Y|D,V}(y \mid d,v)F_V(dv) &= \int 1(g(d,\epsilon) \le y)F_{\epsilon}(d\epsilon) \\
&= \textnormal{Pr}(g(d,\epsilon) \le y) := G(y,d).
\end{align*}
It then follows that the $\tau^{th}$ quantile of $Y_d$ is $G^{-1}(\tau,d)$.

As with the framework of \citen{abadie}, the triangular model under the conditions given above is neither more nor less general than the model of \citen{iqr:ema}.  The key difference between the approaches is that \citen{iqr:ema} uses an essentially unrestricted reduced form but requires monotonicity and a scalar disturbance in the structural equation.  The triangular system on the other hand relies on monotonicity of the reduced form in a scalar disturbance.  In addition, the triangular system, as developed in \citen{imbens:newey}, requires a more stringent independence condition in that the instruments need to be independent of both the structural disturbances and the reduced form disturbance.
That the approaches impose structure on different parts of the model makes them complementary with a researcher's choice between the two being dictated by whether it is more natural to impose restrictions on the structural function or the reduced form in a given application.

The triangular model and the model of \citen{iqr:ema} can be made compatible by imposing the conditions from the triangular model on the reduced form and the conditions from \citen{iqr:ema} on the structural model.  \citen{torgovitsky:id} considers identification and estimation when both sets of conditions are imposed and shows that the requirements on the instruments may be substantially relaxed relative to \citen{iqr:ema} or \citen{imbens:newey} in this case.

\section{Estimation and Inference}

In the previous sections, we have outlined results that are useful for identifying quantile treatment effects and structural functions that are monotonic in a scalar unobservable.  In the following, we briefly review the literature on estimation and inference.  We focus on estimation of the model of \citen{iqr:ema} presented in Section 2 using the moment conditions derived in Theorem 1.  For estimation of the triangular model, see \citen{imbens:newey} for nonparametric estimation and \citen{SLeeTriangularIVQR} for a semiparametric approach.  \citen{abadie} provides results for estimating the QTE for compliers within the LATE-style framework.  Also, we only review approaches for estimating parametric quantile functions: $q(D,X,\tau) = g(D,X,\tau;\theta)$ for $\theta \in \Theta \subset \mathbb{R}^m$.  \citen{HorowitzLeeNonparametricIVQR} and \citen{GagliardiniScaillet} present nonparametric estimation and inference results for the IVQT model using condition (\ref{e1}).

There are two practical issues that make estimation and inference based on condition (\ref{e1}) challenging.  The first is that the sample analog to condition (\ref{e1}) is non-smooth, and the GMM objective function that would be formed by using (\ref{e1}) as the moment conditions is also generically non-convex, even for linear quantile models.  The second problem is that the model may suffer from weak identification as in the standard linear IV model; \citen{stock:survey} provides a useful introductory survey to weak identification and related inference methods in the linear IV model.  In the quantile case, the problem of weak identification is more subtle than in the linear model in that some quantiles may be weakly identified while others may be strongly identified.  The relevant object for defining the strength of identification of a given quantile is the covariance between $D$ and $Z$ weighted by the conditional density function of the unobservable at the given quantile.  See \citen{iqr:joe2} for a formal definition of this object and related discussion.

While the non-smoothness and non-convexity of the GMM criterion complicates optimization, it does not render the approach infeasible, especially when the dimension of $D$ and $X$ is not too large.  \citen{abadie:1997} considered this approach for estimating an income model and provides further discussion.  One could also estimate the model parameters using the Markov Chain Monte Carlo (MCMC) approach of \citen{mcmc}.  This approach bypasses the need for optimization, instead relying on sampling and averaging to estimate model parameters.  Note that this approach is not a cure-all since MCMC requires careful tuning in applications.  It is also worth noting that standard samplers may perform poorly in even simple linear instrumental variables models when identification is not strong; see \citen{vanDijkIVPosteriors}.  In an approach related to optimizing the GMM criterion function directly, \citen{sakata} proposes estimating the parameters of an instrumental variables quantile model by optimizing a different non-smooth, non-convex criterion function.

To partially circumvent the numerical problems in optimizing the full GMM criterion, \citen{iqr:joe} suggest a different procedure termed the inverse quantile regression for the linear quantile model $q(D,X,\tau) = D'\alpha(\tau) + X'\beta(\tau)$.  The basic intuition for the inverse quantile regression comes from the observation that if one knew the true value of the coefficient on $D$, $\alpha(\tau)$, the $\tau^{\textnormal{th}}$ quantile regression of $Y - D'\alpha(\tau)$ onto $X$ and $Z$ would yield zero coefficients on the instruments $Z$.  This observation allows one to effectively concentrate $\beta(\tau)$ out of the problem and leaves a non-smooth, non-convex optimization problem over only the parameters $\alpha(\tau)$.  Since $D$ is low-dimensional in many applications, one can usually solve this optimization problem using highly robust optimization procedures such as a grid-search.

Algorithmically, the inverse quantile regression estimates for a given probability index $\tau$ of interest can be obtained as follows using a grid search over $\alpha(\tau)$:

1. Define a suitable set of values $\{\alpha_j,
j=1,...,J\}$, and estimate the coefficients $\beta(\alpha_j,\tau)$ and $\gamma(\alpha_j,\tau)$ from the model
$Y-D'\alpha_j = X'\beta(\alpha_j,\tau) + Z'\gamma(\alpha_j,\tau) + \epsilon$ by running the ordinary $\tau$-quantile regression of
$Y-D'\alpha_j$ on $X$ and $Z$.  Call the estimated
coefficients $\widehat \beta(\alpha_j, \tau)$ and $\widehat
\gamma(\alpha_j, \tau)$.

2. Save the inverse of the variance-covariance matrix of
$\widehat \gamma(\alpha_j, \tau)$, which is readily available in
any common implementation of the ordinary QR.  Denote this variance-covariance matrix $\widehat A(\alpha_j,\tau)$.
Form $W_n(\alpha_j,\tau) = \widehat\gamma(\alpha_j,\tau)'\widehat A(\alpha_j,\tau)^{-1}\widehat\gamma(\alpha_j,\tau)$. Note $W_n(\alpha_j)$ is the Wald statistic for testing $\gamma(\alpha_j, \tau)=0$.

3. Choose $\widehat \alpha (\tau)$ as a value among
$\{\alpha_j, j=1,...,J\}$ that minimizes $W_n(\alpha,\tau)$. The
estimate of $\beta(\tau)$ is then given by $\widehat
\beta(\widehat \alpha (\tau), \tau)$.

\citen{iqr:joe} and \citen{iqr:joe2} provide conditions under which the resulting estimator for $\alpha(\tau)$ and $\beta(\tau)$ is consistent and asymptotically normal and provide a consistent variance estimator.  \citen{SakataMarmer} provide a similar multi-step algorithm that circumvents the same numeric problems using the objective function of \citen{sakata}.

The good behavior of the asymptotic approximations obtained in \citen{iqr:joe} and \citen{iqr:joe2} rely on strong identification of the model parameters just as in the linear IV case.  Intuitively, strong identification for
a quantile of interest requires that a particular density-weighted covariation matrix between
$D$ and $Z$ is not local to being rank deficient and that the impact of $Z$ is rich enough to guarantee that
the moment equations have a unique solution. The first condition is analogous to the usual full rank condition in linear IV analysis, and the second condition is required because of the nonlinearity of the problem.  Checking these conditions in practice may be difficult, and it is therefore useful to have inference procedures that are robust to violations of these conditions.

Fortunately, there are several inference procedures that remain valid under weak identification.  A nice feature of the algorithm defined for estimating $\alpha(\tau)$ above is that it produces a weak-identification-robust inference procedure naturally as a byproduct.  \citen{iqr:joe2} show that the Wald statistic,  $W_n(\alpha,\tau)$ converges in distribution to $\chi^2_{\dim(Z)}$ under the null that $\alpha = \alpha_0$ where we let $\alpha_0$ denote the true value of $\alpha(\tau)$ without needing either of the conditions discussed in the preceding paragraph.  Thus a valid $(1-p)\%$ confidence region for $\alpha(\tau)$ may be constructed as the set:
 \begin{equation}\label{ci}
 \{ \alpha:  W_n(\alpha,\tau) \leq c_{1-p}\}
 \end{equation}
where $c_{1-p}$ is such that $\textnormal{Pr}(\chi^2_{\dim(Z)} > c_{1-p}) = p$, and the set is approximated numerically by considering $\alpha$'s in the grid $\{\alpha_j, j =1,...,J\}$.  \citen{iqr:joe2} show that confidence region in equation (\ref{ci}) is valid when the model parameters are strongly identified and remains valid when the model is weakly identified or even unidentified.  \citen{SakataMarmer} provide a similar procedure and result for their procedure as well.  \citen{JunWeakIdIVQR} provides yet a different approach to performing weak-identification-robust inference in models defined by conditions (\ref{e1}). Finally, \citen{fsqr} show that one can form statistics for inference about the entire parameter vector $\theta$ that are condtionally pivotal in finite-samples for models defined by quantile restrictions such as (\ref{e1}).  Since the statistics do not depend on unknown nuisance parameters in finite samples, the exact distributions of these statistics can be calculated and inference can proceed without relying on asymptotic approximations or statements about the strength of identification.  The distributions produced in \citen{fsqr} are not standard and so must be calculated by simulation.

\section{Conclusion and Directions for Future Research}

In this paper, we have reviewed approaches for building quantile models in the presence of endogeneity, focusing on conditions that can be used for identification.  We have also briefly reviewed some of the practical issues that arise in estimation of instrumental variables quantile models and approaches to dealing with these issues.  The models and estimation strategies outlined and cited in this review have already seen use in empirical economics where they have mostly been used for their ability to uncover interesting distributional effects.  In this review, we have also noted that the identification strategy employed in this paper can be used to uncover structural objects even if quantile effects are not the chief objects of interest as in \citen{BerryHaileDiscreteChoiceId}.

While the results reviewed in this paper are useful in a variety of contexts, there remain interesting areas for research in quantile models with endogeneity.  In some applications, features of the conditional distribution are not the chief objects of interest and researchers are interested in effects of treatments on unconditional quantiles.  Given the set of conditional quantiles, such unconditional effects may be uncovered.  In recent work, \citen{FroelichMellyUnconditionalIVQR} propose a different approach, related to \citen{abadie}, to estimating structural effects of endogenous variables on unconditional quantiles directly. It would also be interesting to think about quantile-like quantities for multivariate outcomes with endogenous covariates.   The results reviewed in this paper offer one possible approach for quantile modeling with endogeneity, but there remain many interesting directions and other approaches to be explored in further research.

\appendix

\section{Proofs}

\subsection{Proof of Theorem 1}
 Conditioning on $X=x$ is suppressed. For $P$ almost every value $z$ of $Z$,
 \bsnumber\label{pr1_1}
P\[ U_{\sss D} \leq  \tau | Z=z\] & \overset{(1)}= \int P\[ U_{\sss D} \leq  \tau | Z=z, V =v \] d
 P\[V=v |Z=z \] \\
 &\overset{(2)}= \int P\[ U_{ \delta(z, v)} \leq  \tau | Z=z, V =v \] d
 P\[V=v |Z=z \] \\
 &\overset{(3)}= \int P\[ U_{0} \leq  \tau | Z=z, V =v \] d
 P\[V=v |Z=z \] \\
 &\overset{(4)} = P\[U_{0} \leq \tau |Z=z \] \overset{(5)}= \tau.
\end{split}\end{align}
Equality (1) is by definition. Equality (2) is by the
representation A3. Equality (3) is by the similarity assumption
A4 and representation A3: Conditional on $(V=v,Z=z)$,
$D=\delta(z,v)$ is a constant, so that by A4, $U_{\delta(z,v)} \text{ has the same distribution as  }
U_{0},$  where $``0"$ denotes any fixed value of $D$. Equality (4) is by
definition, and equality (5) is by the independence assumption A2.
This shows the first result.

The second result follows from the first and the equivalence of the events
$ \{  q(D, U) \leq q(D, \tau) \}  = \{ U \leq
\tau\}$ under $u \mapsto q(d,u)$ strictly increasing for each $d$ on the domain $[0,1]$. To show
the third result we note that
$$
\{ U  \in I \} \subseteq \{  \{u: q(D,u) = q(D,U)\} \cap I \neq \oslash \}.
$$
Since $Y =q (D,U)$, the latter event is equivalent to the event $\{Y \in q(D,I)\}$, where $q(D,I)$ denotes
the image of $I$ under the mapping $u \mapsto q(D,u)$. The third result then follows from
the first result.  \qed

\noindent \subsection{Proof of Theorems \ref{theorem: binary id} and \ref{theorem:discrete D}.} The local identification results follow by a standard argument,
introduced in \citeasnoun{rothenberg:1971}, which we omit for brevity. The global identification result is obtained as follows. By assumption  $q
\in \mathcal{L}$. Hence, we need to check whether $y=q$ is the only
solution to $\Pi(y)=0$ over $\mathcal{L}$. Consider a covering set
$\mathcal{L}_j$ and the $l$-permutation $m(j)$ corresponding to it,
as defined in the theorem. By assumption $ \Pi_{m(j)}(q)
= 0$. By assumption $q \in \mathcal{L}_j$. The
stated rank conditions, compactness, and convexity of
the polytope $\mathcal{L}_j$ imply that the mapping $y \to \Pi_{m(j)}(y)$,
which maps $\mathcal{L}_j \subset \mathbb{R}^l$ to $\mathbb{R}^l$, is a
homeomorphism (one-to-one) between $\mathcal{L}_j$ and $\Pi_{m(j)}
(\mathcal{L}_j) $ by the global univalence theorem, Theorem 1 of
\citen{mas-colell:convex}. Thus, $y=q$ is the unique solution of $\Pi_{m(j)}(y) = 0 $ over
$\mathcal{L}_j$. Since this argument applies to every $j$ and
$\{\mathcal{L}_j\}$  cover $\mathcal{L}$, it follows that $y=q$ is
the unique solution of $\Pi(y) = 0 $ over $\mathcal{L}$. \qed

\subsection{Proof of Theorem \ref{theorem: general id}.} We have that $q$ solves
$ P[ Y \leq q(D) |Z] = \tau$ a.s., and $q \in
\mathcal{L}$ by assumption. Hence we need to check whether $q$ is the only
solution to $P[ Y \leq q(D) |Z] = \tau$ a.s. in $\mathcal{L}$.
Suppose there is $m \in \mathcal L$ such that $P[Y \leq m(D) |Z]=
\tau\ \text{ a.s.}$ Define $\Delta(d) := m(d)-q(d)$, and
write \bsnumber P[Y \leq m(D) |Z] & - P[Y \leq q(D) |Z]
\overset{(1)}=
  E [ E [  \int_0^1
f_{\epsilon}(\delta \Delta(D) |D, Z)  \Delta (D) d \delta |D, Z]
|Z ]   \\
 & \overset{(2)} =    E [  \int_0^1 f_{\epsilon}(\delta \Delta(D)
|D, Z)  \Delta (D) d \delta  |Z ] \\  &\overset{(3)} =
 E [ \Delta (D) \cdot \omega_{\Delta}(D,Z) |Z].\label{idp21}
\end{split}\end{align}
Noting that  (1) follows by the fundamental
 theorem of calculus, (2) by the
law of iterated expectations, and (3) by linearity of the Lebesgue
integral. For uniqueness we need that (\ref{idp21})=0 a.s.
$\Rightarrow$ $\Delta(D) = 0$ a.s., which is assumed.  The result (i) follows.

Result (ii) is immediate from (i) by  the triangle inequality for $\|\cdot\|_{p,P}$.
\qed

\subsection{Proof of Sufficiency of (\ref{LMR}) and (\ref{POS})} Here show that (\ref{LMR}) and (\ref{POS}) are sufficient  for identification over the parameter space $\mathcal{L} = (q+ C) \cap H$.   Note that in general $\mathcal{L}$ has at most up to five
edges: the left and the right edges, parallel to each other, the top and the bottom edges,
 also parallel to each other, and the edge generated by the intersection of $\mathcal{L}$ with the $45$ degree line.  Let $e_1$ and $e_2$ be coordinate vectors in $\mathbb{R}^2$, and let $L_k$ denote the various subspaces spanned by faces of $\mathcal{L}$  containing $y$. In particular, we have that $L_1 := \mathbb{R}^2$ for all $y$ in the two-dimensional face $F_1:= \mathcal{L}$, $L_2 := \textrm{span}(e_2)$ for all $y$ in the one-dimensional faces given by the left and the right edge of $\mathcal{L}$,  denoted both by $F_2$,   $L_3 := \textrm{span}(e_1)$ for all $y$ in the two-dimensional faces given by the top and bottom edges of $\mathcal{L}$, denoted both by $F_3$,   and  $L_4 = \textrm{span}(e_1+e_2)$ for all $y$ in the one-dimensional face $F_4$ given by the edge generated by the intersection of $\mathcal{L}$ with the 45 degree line.  The subspaces spanned by vertices, which are zero-dimensional faces of $\mathcal{L}$, are null spaces; so we do not need to consider them.  We compute the projections of the Jacobian map onto these subspaces: $\text{proj}_{L_1} \circ \partial \Pi(y)[l] = \partial \Pi(y)l$, $\text{proj}_{L_2}\circ \partial \Pi(y)[l] =   f_{\sss Y,D}(y_0,0|Z=0) l$, $\text{proj}_{L_3} \circ \partial \Pi(y)[l] =  f_{\sss Y,D}(y_1,1|Z=1) l $, $\text{proj}_{L_4} \circ \partial \Pi(y)[l] =   \{[f_{\sss Y,D}(y_1,1|Z=1)  + f_{\sss Y, D}(y_0, 0|Z=1)  + f_{\sss Y,D}(y_1,1|Z=0) +  f_{\sss Y,D}(y_0,0|Z=0)]/2 \} l$, for $y \in F_k$ and $l \in L_k$ in each of the cases. We then compute the corresponding determinants of the maps
 $$
 \text{proj}_{L_k}\circ \partial \Pi(y): L_k \to L_k,
 $$
where determinants are computed with respect to the coordinate systems of $L_k$, as $\det [\partial \Pi(y)]$ for $k=1$, $ f_{\sss Y,D}(y_0,0|Z=0)$ for $k=2$, $f_{\sss Y,D}(y_1,1|Z=1)$ for $k=3$, $[f_{\sss Y,D}(y_1,1|Z=1)  + f_{\sss Y, D}(y_0, 0|Z=1)  + f_{\sss Y,D}(y_1,1|Z=0) +  f_{\sss Y,D}(y_0,0|Z=0)]/2$ for $k=4$. Theorem 2 requires that these determinants are positive for values of $y \in F_k$.  This condition  is implied by the simpler conditions (\ref{LMR}) and (\ref{POS}).  For the case of $\mathcal{L} =q+ C$, verification is analogous except that we do not need to consider $L_4$.  Thus, the positive determinant condition of Theorem 2 is implied by the conditions (\ref{LMR}) and (\ref{POS}) for $\mathcal{L} = q+C$ as well. \qed

\bibliographystyle{econometrica}

\bibliography{mybibIVQRReview}

\end{document}